# Enhancing Forecasting Accuracy in Dynamic Environments via PELT-Driven Drift Detection and Model Adaptation

Nikhil Pawar[1], Guilherme Vieira Hollweg[2], Akhtar Hussain[3], Wencong Su[2,*], and Van-Hai Bui[2,*]

[1]Department of Computer and Information Science, University of Michigan—Dearborn, Dearborn, MI 48128, USA; inikhil@umich.edu (N.P.)

[2]Department of Electrical and Computer Engineering, University of Michigan—Dearborn, Dearborn, MI 48128, USA

[3]Department of Electrical and Computer Engineering, Laval University, Quebec City, QC G1V 0A6, Canada

*Correspondence: vhbui@umich.edu (V.H.B) and wencong@umich.edu (W.S.)

***Abstract:*** **Accurate time series forecasting models are often compromised by data drift, where underlying data distributions change over time, leading to significant declines in prediction performance. To address this challenge, this study proposes an adaptive forecasting framework that integrates drift detection with targeted model retraining to compensate for drift effects. The framework utilizes the Pruned Exact Linear Time (PELT) algorithm to identify drift points within the feature space of time series data. Once drift intervals are detected, selective retraining is applied to prediction models using Multilayer Perceptron (MLP) and Lasso Regressor architectures, allowing the models to adjust to changing data patterns. The effectiveness of the proposed approach is demonstrated on two datasets: a real-world dataset containing electricity consumption and HVAC system data, and a synthetic financial dataset designed to test cross-domain applicability. Initial baseline models were developed without drift detection using extensive feature engineering. After integrating drift-aware retraining, the MLP model achieved a 44% reduction in mean absolute error (MAE) and a 39% increase in $R^2$ on the real-world dataset, while even greater improvements were observed on the synthetic financial dataset. Similar enhancements were achieved with the Lasso Regressor. These results highlight the robustness and generalizability of incorporating drift detection and adaptive retraining to sustain forecasting accuracy across diverse domains.**

## 1  Introduction

Time series forecasting plays a critical role across a wide range of domains such as energy consumption modeling, HVAC control, stock market prediction, and policy planning. Accurate forecasts drive efficient decision-making, optimize resource use, and reduce operational costs. However, real-world time series data often experience dynamic structural changes over time, making forecasting particularly challenging. These changes, known as concept drift, refer to the shifting statistical properties of the data-generating process, which can undermine the reliability of predictive models if not explicitly addressed [1]–[5]. As time series data evolve, the relationship between predictors and target variables may change due to external influences, structural shifts, or latent temporal factors. For instance, consumer behavior patterns in electricity usage can change during seasonal transitions, while interest rate dynamics can be influenced by macroeconomic shocks. Such changes violate the fundamental assumption of stationarity that many forecasting models depend on. This leads to significant degradation in predictive performance over time [6]–[9]. Concept drift can manifest in various forms: abrupt, gradual, incremental, or recurring. Hence, developing models that are robust to these temporal fluctuations has become a pressing concern.

Existing strategies for handling drift in machine learning and time series forecasting can be broadly classified into two categories: drift detection and model adaptation. On the detection side, researchers have explored approaches that monitor statistical properties of the input data to flag changes, such as the Pruned Exact Linear Time (PELT) algorithm, which identifies change points in time series with a guaranteed optimal segmentation [2,10]. On the adaptation side, frameworks like Proceed proactively adapt forecasting models by comparing recent training and test samples to anticipate drift [1]. Similarly, dynamic models like DA-LSTM adjust their structure to accommodate newly emerging patterns without requiring explicit drift boundaries [3]. Recent advances in distributional adaptation have led to techniques like the Temporal Conditional Variational Autoencoder (TCVAE), which captures evolving multivariate dependencies through latent temporal embeddings [5]. These methods reflect a growing consensus that fixed-window training and naive retraining are insufficient for real-world applications. However, they often require complex architectures or extensive domain-specific tuning, which limits their generalizability [6]. Our research builds on these insights and seeks to design a general-purpose, interpretable, and efficient framework capable of handling drift through more targeted model adjustments.

In parallel, the challenge of benchmarking drift-aware forecasting models has also gained attention, particularly given that many evaluations are conducted on synthetic or oversimplified datasets that fail to reflect the complexity of real-world environments [7]. For drift-aware forecasting to be practical, models must adapt across domains, efficiently managing real-world noise and data shifts without resorting to full model retraining at every update. While several surveys have reviewed drift detection and ensemble learning techniques [8,9], limited work has systematically compared baseline and retrained models using actual drift timestamps to guide retraining. Our work addresses this gap by proposing a lightweight and interpretable solution that leverages robust change detection via PELT, enabling adaptive retraining with minimal domain-specific tuning [13-15]. In doing so, we integrate changepoint-based drift detection with selective model retraining, offering a unified framework that combines both detection and adaptation in a practical, efficient manner.

To validate our approach, we develop an adaptive time series forecasting pipeline that employs PELT to detect change points within feature distributions and triggers retraining only for these drift-affected intervals.. This strategy avoids continuous retraining and reduces unnecessary computation while directly addressing the portions of the dataset that exhibit drift. Compared to prior approaches that either fully retrain or use sliding windows, our selective retraining is localized and efficient. For experimentation, we use two types of datasets. First, a real-world dataset containing time series of electricity consumption and HVAC operation data—domains known for exhibiting natural drift due to seasonal, operational, and occupancy-related fluctuations. Second, we generate a synthetic financial dataset that includes abrupt drift events embedded into interest rate patterns. This dataset helps assess our method's cross-domain generalizability, which is rarely tested in existing studies [10,11]. We evaluate two model families under our framework: Multilayer Perceptron (MLP) for capturing non-linear dependencies and Lasso Regressors for linear and interpretable modeling. The baseline models are trained on the full dataset without any drift handling, using standard feature engineering techniques. The PELT-retrain models, in contrast, use detected change point intervals as training blocks, allowing the model to re-learn only when meaningful distributional shifts occur. This approach ensures the retrained model remains current without forgetting stable regions [12]. Our results demonstrate significant improvements. In the real-world dataset, the PELT-aware MLP achieves up to a 44% reduction in MAE and a 39% increase in R², compared to its baseline counterpart. The synthetic dataset shows even more substantial gains, validating the model's drift-handling capability across different domains. The Lasso model follows a similar trend, though with more modest improvements, illustrating that even simpler models benefit from targeted retraining.

## 2  Related Work

Adaptive forecasting in the presence of concept drift has been explored through various architectures, particularly within domains like smart grids and financial systems. Notable recent methods include DA-LSTM [1], which dynamically adjusts LSTM weights during forecasting, and TCVAE [2], which utilizes latent-space reweighting via a temporal variational autoencoder. While both techniques demonstrate strong adaptability, their computational demands, complexity, and limited interpretability restrict deployment in lightweight or embedded environments.

In contrast, our approach introduces a modular and resource-efficient pipeline: drift events are explicitly detected using the Pruned Exact Linear Time (PELT) algorithm, and only a compact model is retrained on drift-affected segments, reducing overhead while preserving adaptability. This makes it particularly suitable for embedded deployments in electricity grid systems or financial platforms.

Table1 highlights a direct comparison between these models, evaluating key dimensions such as model complexity, retraining logic, domain generalizability, and suitability for low-power systems. As seen, our PELT-triggered retraining strategy offers a favorable balance of efficiency, interpretability, and domain flexibility, without compromising adaptability to sudden shifts.

| Criteria | DA-LSTM | TCVAE | Our PELT-Retrain |
| --- | --- | --- | --- |
| **Model Type** | LSTM | Temporal VAE | MLP, Lasso |
| **Drift Handling** | Online updates | Latent reweighting | PELT-based detection |
| **Retraining Strategy** | Continuous | Global offline | Selective local |
| **Computational Cost** | High | Very High | Low |
| **Interpretability** | Low | Low | High |
| **Deployment Fit** | Needs tuning | High memory footprint | Lightweight, modular |
| **Embedded Readiness** | Not ideal | Not suitable | Suitable |
| **Drift Trigger** | Implicit (loss-based) | Implicit (latent drift) | Explicit (PELT changepoints) |
| **Sudden Drift Response** | Moderate | High | High |

Table 1: Comparison of Drift Adaptation Strategies

This contrast motivates our focus on explicitly triggered, low-cost retraining pipelines that scale across application domains with minimal parameter tuning.

# 3 Proposed Framework for Data Drift Detection

## 3.1 Data Drift Formulation

Data drift refers to a shift in the statistical properties of input features over time, often leading to model degradation if not accounted for. This phenomenon is particularly critical in time series forecasting, where changing patterns may directly impact predictive accuracy.

Two major forms of drift are:

- Covariate Shift, where the distribution of the input features changes, but conditional output remains stable:

$$P_t(X) \neq P_{t+1}(X), \text{ while } P_t(Y|X) = P_{t+1}(Y|X) \tag{1}$$

- Concept Drift, where the mapping from input to output itself evolves:

$$P_t(Y|X) \neq P_{t+1}(Y|X) \tag{2}$$

We model the problem as a change point detection task, where drift points $\tau_1, \tau_2, \ldots, \tau_m$ are timestamps at which the data distribution changes significantly:

$$T = \{\tau_1, \tau_2, \ldots, \tau_m\}, where \ P_{\tau_i}(X) \neq P_{\tau_{i+1}}(X) \tag{3}$$

Given a time series $y_{1:n}$, our objective is to partition the sequence into $m+1$ segments with homogeneous statistical behavior:

This formulation allows us to detect both sudden and gradual changes across a variety of time series domains without relying on the target variable, enabling fully unsupervised drift detection on input features.

## 3.2 Proposed Detection Framework

Our proposed detection framework is centered on the Pruned Exact Linear Time (PELT) algorithm, chosen for its efficiency and accuracy in identifying drift points in time series. PELT segments data into blocks with stable statistical properties, allowing us to isolate regions where meaningful distributional changes occur in input features. This algorithm was introduced by Killick et al. for efficient and exact multiple changepoint detection in O(n) time [16]. These drift timestamps then inform our retraining strategy.

Let the time series be denoted as:

$$y_{1:n} = (y_1, y_2, \ldots, y_n) \tag{4}$$

The goal of PELT is to identify change points $\tau_1, \ldots, \tau_m$ that divide the series into $m+1$ segments:

$$0 = \tau_0 < \tau_1 < \cdots < \tau_m < \tau_{m+1} = n$$

To achieve this, PELT minimizes the following objective:

$$\min \left( \sum_{i=1}^{m+1} C\left(y_{(\tau_{i-1}+1):\tau_i}\right) + \beta f(m) \right) \tag{5}$$

where $C(y_{(\tau_{i-1}+1):\tau_i})$ is the cost of segment, and $\beta f(m)$ is a penalty function that controls the number of change points detected, typically with $f(m) = m$ to apply a linear penalty.

The cost of each segment is computed based on its likelihood. A general form of the segment cost is defined as the negative log-likelihood:

$$C(y_{\tau_{i-1}+1:\tau_i}) = -\max_{\theta} \sum_{j=\tau_{i-1}+1}^{\tau_i} \log f(y_i|\theta) \qquad (6)$$

where $f(y_j|\theta)$ is the assumed data distribution and $\theta$ are the estimated parameters for that segment.

For continuous, normally distributed data, the cost simplifies to the least squares error:

$$C(y_{t_1:t_2}) = \sum_{t=t_1}^{t_2} (y_t - \bar{y}_{t_1:t_2})^2 \qquad (7)$$

Where $\bar{y}_{t_1:t_2}$ is the mean of the segment $[t_1, t_2]$.

PELT improves efficiency using a pruning condition that eliminates suboptimal change point candidates. This condition is given by:

$$F(t) + C(y_{t+1:s}) + K \geq F(s) \qquad (8)$$

This pruning step reduces the computational load, allowing PELT to achieve expected linear time complexity $O(n)$ under realistic data conditions, while still guaranteeing globally optimal change point segmentation.

### 3.3 Visual Patterns of Data Drift

Having formalized the concept of data drift and introduced our detection framework based on the PELT algorithm, we now emphasize the importance of visualizing drift in the prediction target itself. While the changepoint detection in our pipeline operates on engineered features, plotting the temporal evolution of the target columns (elec_kW) in the real-world power dataset and (interest_rate) in the synthetic financial dataset offers direct and interpretable evidence of drift in the quantity being forecasted. This approach is both valid and essential. The target variable encapsulates the real-world signal our models aim to predict, and observable changes in its behavior over time reflect underlying shifts in data distribution. Such shifts alter the mapping from inputs to outputs i.e., from f(x)→y and thereby justify the need for model retraining. Visualizing these changes serves as a diagnostic and motivational tool, demonstrating why static models often fail in non-stationary environments. A model trained on historical data cannot reliably track a target that has drifted, unless the learning system itself adapts [17].

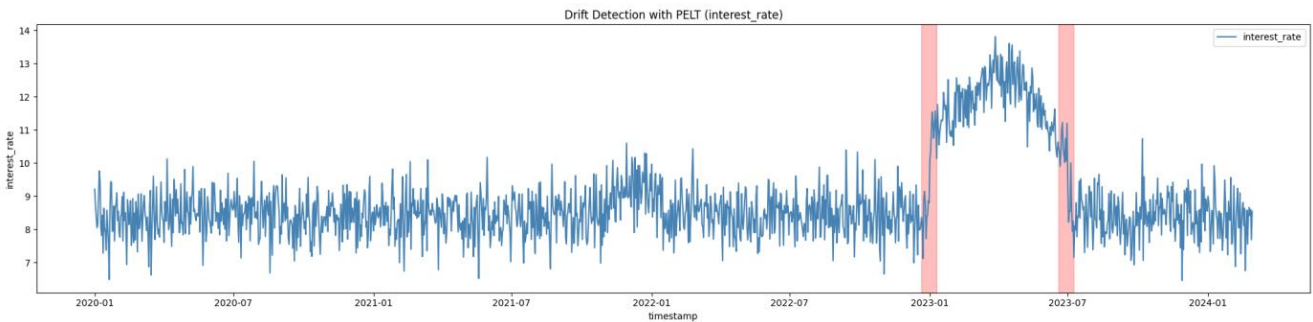

**Figure 1: Drift Detection with PELT (interest_rate)**

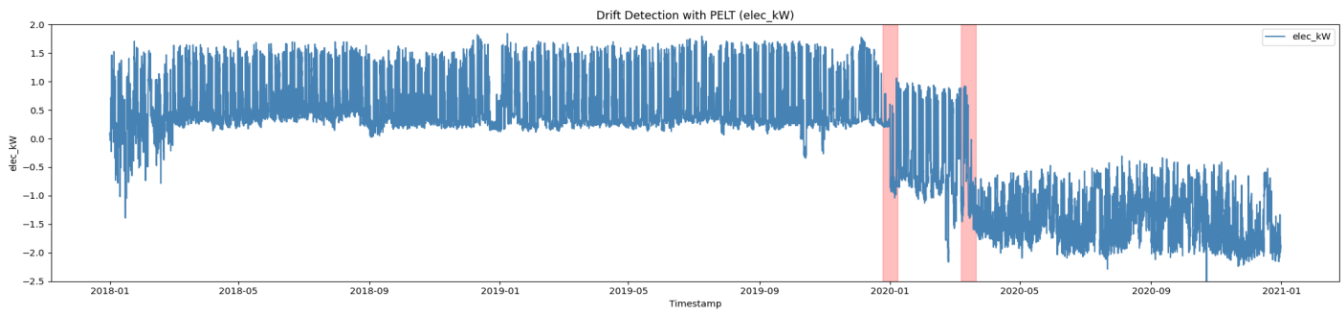

**Figure 2: Drift Detection with PELT (elec_kW)**

The strength of our detection framework is underscored by its alignment with these visual patterns. In the synthetic dataset, Figure1 illustrates how the PELT algorithm accurately identifies changepoints corresponding to the shifts in interest_rate, validating the framework's responsiveness to drift. Similarly, in the real-world dataset, Figure2 reveals seasonal changes in elec_kW that are correctly flagged by the algorithm confirming its utility in uncovering real-world drift scenarios. Together, these visuals reinforce the interpretability and effectiveness of our pipeline's drift detection component.

## 4 Forecasting Framework Design

To evaluate impact of data drift and assess the effectiveness of localized retraining strategies, we designed two forecasting pipelines, a baseline model that assumes data stationarity and uses the entire training dataset, and a drift-aware retraining model that selectively adapts to distributional shifts using changepoints detected via the PELT algorithm. While both pipelines share the same underlying feature engineering strategy and support two model types MLP and Lasso, they differ fundamentally in how they treat temporal dynamics in the training data.

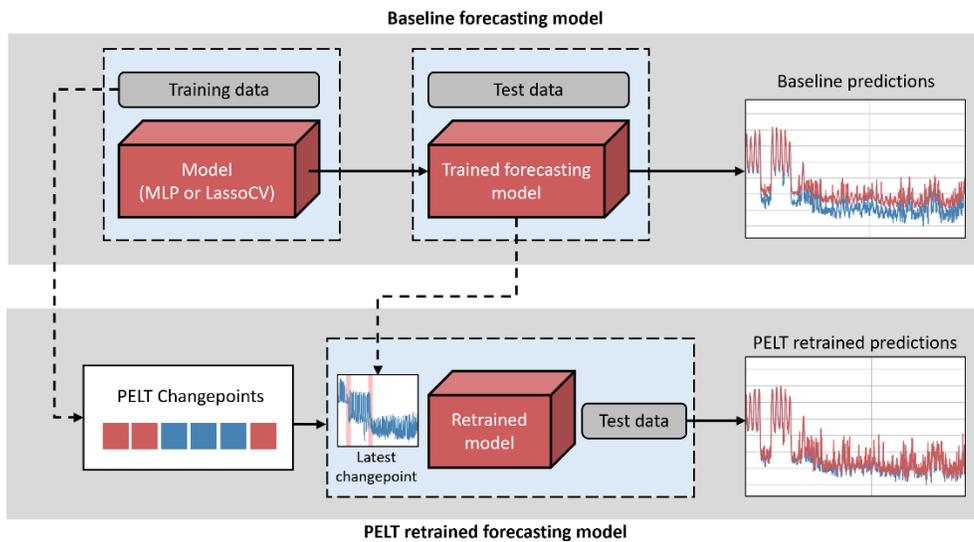

**Figure 3: Forecasting Framework overview diagram**

The overall goal of this forecasting framework is to isolate and quantify the advantages of integrating drift detection into the time series prediction, especially under evolving data distributions. The modular design makes this framework applicable across different domains, whether dealing with real-world consumption data or synthetic financial signals. Similar selective retraining strategies have shown performance benefits in traffic forecasting scenarios [18], while broader literature in concept drift detection highlights the value of performance-aware retraining triggers like changepoint detection [19]. In the subsections below, we detail each pipeline using high-level diagrams to clearly illustrate their conceptual flow and logic.

## 4.1 Baseline Forecasting Strategy

The baseline forecasting strategy adopts a static learning paradigm, where a model is trained once using entire available training period, with the assumption that past data remains consistently representative of the future. No drift detection or model update is performed. This traditional approach provides a useful reference point to assess how much performance deteriorates when no adaptation is applied under changing data conditions. As depicted in Figure3, the full training dataset is first processed and directly used to train a forecasting model, either an MLP neural network or a Lasso regressor.

Once the model is trained, it is evaluated on a separate, unseen test set. Since the model is never updated after its initial training, this setup helps identify whether the model suffers from distributional mismatch in dynamic environments [19]. Key characteristics of the baseline framework include the use of the entire historical data for training, the absence of any drift detection mechanism, and reliance on a single static model for forecasting. Evaluation is performed on future data without any retraining or fine-tuning steps. This approach is simple and computationally efficient but may underperform when the input data exhibits non-stationary behavior.

## 4.2 Drift-Aware Retraining Strategy

The drift-aware forecasting strategy introduces adaptability into the modeling process by selectively retraining the model only on the segment of data that follows the most recently detected changepoint. To identify these changepoints, the PELT algorithm is applied to the input features within the training set, allowing the model to recognize potential distributional shifts that may affect future performance. Once the last drift point is detected, the training data prior to this point is discarded, and the model is retrained from scratch using only the post-drift segment. This enables the model to focus on the most recent patterns and avoid outdated trends that might negatively influence generalization. As illustrated in Figure3, the forecasting pipeline begins with the full historical training dataset, from which the most relevant segment is dynamically extracted based on drift detection. The retraining phase uses the same modeling architecture as the baseline, either a Multi-Layer Perceptron (MLP) or Lasso regressor but operates only on the updated data segment. Like prior studies, such selective retraining based on recent patterns improves forecast relevance under evolving distributions [18]. The retrained model is then evaluated on a fixed future test set, identical to the one used in the baseline strategy, to ensure consistency and comparability between the two forecasting approaches.

This retraining approach is designed to strike a balance between accuracy and computational efficiency. By using only the most recent post-drift data, it increases the relevance of the training inputs while avoiding the complexity of retraining after every detected changepoint or using arbitrary time windows. This design choice is particularly beneficial for applications where retraining must remain lightweight, yet responsive to evolving data regimes. It also ensures the method remains flexible and generalizable across various domains without overfitting to transient shifts. The effectiveness of this strategy is later evaluated in the results section, where it is compared against the static baseline model.

## 5 Experimental Setup

This section outlines the experimental design used to evaluate the effectiveness of our proposed drift-aware retraining framework. We describe the datasets utilized, the model configurations for both baseline and adaptive strategies, and the evaluation metrics employed to assess forecasting performance under real and synthetic drift conditions. Our setup is structured to highlight how drift-aware retraining improves model robustness and accuracy compared to static models in dynamic environments.

### 5.1 Datasets

To evaluate the robustness and generalizability of our drift-aware forecasting framework, we employ two distinct datasets: a real-world building energy consumption dataset and a synthetically generated financial dataset. These datasets differ not only in domain and data-generating processes but also in the type and manifestation of data drift, enabling us to validate our approach under both natural and induced drift conditions.

The real-world dataset used in this study is obtained from Building 59 at the Lawrence Berkeley National Laboratory (LBNL), a medium-sized commercial office building equipped with panel-level submeters. Spanning from January 1, 2018, to December 31, 2020, this dataset comprises 26,287 hourly readings capturing electricity demand across diverse subsystems including plug loads, lighting, and HVAC. These measurements inherently reflect fluctuations driven by occupancy schedules, equipment usage, seasonal temperature shifts, and external events, making it an ideal candidate for investigating real-world drift in time series forecasting. In the context of our research, which emphasizes localized retraining triggered by drift detection, this dataset provides a realistic setting to test our framework's ability to isolate and adapt to such dynamic changes. Data preprocessing involved timestamp normalization and forward-filling for missing values to preserve temporal consistency. A rolling average smoothing step was applied solely for visualization and interpretability during analysis, no leakage of this smoothing was introduced into model training. This dataset's naturally embedded drift patterns challenge static models and underscore the necessity of selective retraining mechanisms like those developed in our pipeline, where the PELT algorithm identifies significant distributional shifts that guide efficient model updates.

To complement the real-world data, a synthetic dataset was generated to simulate controlled drift scenarios in a financial forecasting context. This dataset spans from January 1, 2020, to December 31, 2023, with hourly observations representing synthetic interest rate values. The data was designed to include both sudden and gradual drifts, enabling precise evaluation of our PELT-based detection and localized retraining strategy. For instance, abrupt increases in interest rates were introduced in early 2023 to mimic macroeconomic shocks such as policy changes, while slower, progressive shifts were embedded to simulate economic cycles and long-term trends. The synthetic nature of this dataset offers full control over drift injection, making it an effective tool for benchmarking the sensitivity and adaptability of our framework. All drift changes were confined to feature space, ensuring that model performance is assessed under realistic but controlled distributional shifts. Preprocessing steps included timestamp alignment and basic outlier filtering, ensuring the synthetic series retained a clean, interpretable structure without artificial smoothing that could mask change points. This dataset plays a crucial role in validating the core premise of our work, namely that selective retraining based on drift detection can offer computationally efficient yet accurate alternative to full retraining in dynamic environments.

## 5.2 Feature Engineering

To ensure the predictive models captured both temporal and contextual dynamics, we performed comprehensive feature engineering tailored to each dataset. This involved time-based cyclic encodings such as hour_sin and hour_cos, which transform the hour of the day into continuous signals using sine and cosine functions, allowing models to learn daily periodic patterns. Similarly, dow_sin and dow_cos encode the day of the week to help capture weekly consumption cycles. Lag features like lag_1, lag_24, and lag_168 provide access to electricity usage in the past 1, 24, and 168 hours, respectively, enabling the model to learn from short- and long-term dependencies. Rolling statistics, including roll_mean and roll_std, summarize recent consumption behavior over 24-hour and 168-hour windows, helping smooth out noise and highlight trend shifts. Table2 summarizes all features used in both datasets and their roles within the modeling pipeline.

| Feature Name | Type | Purpose / Description |
| --- | --- | --- |
| hour_sin, hour_cos | Time encoding | Capture daily periodicity |
| dow_sin, dow_cos | Time encoding | Encode weekly cyclic behavior |
| lag_1, lag_24, lag_168 | Lag features | Capture short-, daily-, and weekly-term dependencies |
| roll_mean_24, roll_std_24 | Rolling stats (24h) | Represent recent consumption trends and variability |
| roll_mean_168, roll_std_168 | Rolling stats (7d) | Capture weekly trend dynamics and detect longer-term fluctuations |

Table 2: Summary of engineered features and their respective roles in the predictive pipeline.

## 5.3 Model Configuration

To maintain methodological consistency and ensure fair cross-dataset comparisons, we employ two core model types—Multi-Layer Perceptron (MLP) and Lasso Regression—across both the real-world and synthetic datasets. Each model is evaluated under two strategies: a baseline model trained on the full dataset and a drift-aware retraining model trained on post-drift segments identified via the PELT algorithm. This symmetric structure enables a controlled investigation of how drift-aware adaptation impacts forecasting performance across different model complexities and data characteristics. The MLP model is implemented as a simple feedforward neural network consisting of two hidden layers. Each hidden layer contains 64 neurons and uses the ReLU activation function, followed by a dropout layer with a rate of 0.2 to reduce overfitting. The final output layer has a single neuron for regression. We train the model using the Adam optimizer with a learning rate of 0.001, mean squared error (MSE) as the loss function, and early stopping based on validation loss. The batch size is fixed at 64, and training is capped at a maximum of 300 epochs, with early termination if the validation loss does not improve for 10 consecutive epochs.

For the Lasso regression model, we employed the LassoCV implementation from scikit-learn, which performs automatic hyperparameter tuning via time series cross-validation. Instead of manually selecting the regularization strength (alpha), we provided a candidate set of values [0.001, 0.01, 0.1, 1.0], and the optimal value was selected using a 5-fold TimeSeriesSplit to ensure temporal integrity and prevent look-ahead bias. This approach improves generalization while maintaining model interpretability and leveraging the inherent sparsity-promoting nature of L1 regularization. In the baseline scenario, the model is trained on the entire pre-drift dataset using a reduced set of causally engineered features such as rolling means, Fourier seasonality components, and time-of-day indicators. For the retrained scenario, we restrict the training data to the post-drift segment identified using the PELT algorithm and enrich the feature set with additional lag-based signals, calendar features, and trend-related components. This separation ensures that retraining decisions are based on distributional shifts rather than architectural changes, allowing for a fair and controlled evaluation of the drift-aware methodology. Both pipelines use standardized input features, and the same LassoCV configuration is preserved across scenarios to isolate the impact of retraining. This uniform treatment ensures that any observed performance gains can be confidently attributed to drift-aware retraining rather than differences in model complexity or tuning strategies.

## 5.4 Evaluation Metrics

To evaluate the accuracy and robustness of our forecasting models under both stationary and non-stationary conditions, we employ three widely accepted regression evaluation metrics: Mean Absolute Error (MAE), Root Mean Squared Error (RMSE), and Coefficient of Determination (R²). These metrics were selected to capture different aspects of prediction error and offer a comprehensive assessment of model performance [20].

### 5.4.1 Mean Absolute Error (MAE)

$$MAE = \frac{1}{n}\sum_{i=1}^{n} |y_i - \hat{y}_i| \qquad (9)$$

MAE measures the average magnitude of errors between predicted and actual values, without considering their direction. It provides an intuitive interpretation of the typical absolute error in the same units as the target variable. In our experiments, MAE helps quantify overall deviation without disproportionately emphasizing outliers, making it particularly useful for evaluating stable models across gradual shifts in data distribution.

### 5.4.2 Root Mean Squared Error (RMSE)

$$\text{RMSE} = \sqrt{\frac{1}{n}\sum_{i=1}^{n}(y_i - \hat{y}_i)^2} \qquad (10)$$

RMSE is the square root of the average squared differences between predictions and actual values. Unlike MAE, RMSE penalizes larger errors more severely, thus offering a more sensitive measure of performance when significant deviations occur. We use RMSE to assess model resilience in the presence of abrupt drifts and rare but impactful deviations, especially in synthetic datasets where large changes were intentionally introduced.

### 5.4.3 Coefficient of Determination (R²)

$$R^2 = 1 - \sum(y_i - \hat{y}_i)^2 / \sum(y_i - \bar{y})^2 \qquad (11)$$

R² measures the proportion of variance in the target variable that is explained by the model. An R² value close to 1.0 indicates high predictive power, while negative values suggest the model performs worse than a simple mean-based predictor. In this study, $R^2$ complements MAE and RMSE by offering a normalized measure of goodness-of-fit, which is useful for comparing model effectiveness across both real-world and synthetic datasets.

These metrics were chosen to ensure a balanced evaluation across both global and localized prediction quality. MAE provides robustness against outliers and offers straightforward interpretability. RMSE emphasizes large errors, which are critical in drift-prone environments. $R^2$ offers insight into model generalization and the effectiveness of learning temporal dynamics. Together, this trio of metrics ensures that our analysis captures both the average and extreme performance characteristics of baseline and retrained models under various drift conditions. Their complementary nature makes them particularly suitable for our goal of understanding the benefits of drift-aware retraining in time series forecasting.

## 6 Results and Analysis

This section presents a comparative analysis of forecasting performance across baseline and drift-aware retrained models. We assess both the MLP and LassoCV on two datasets: the real-world electricity dataset and the synthetically generated financial dataset. In the discussion that follows, we analyze trends, improvements, and insights derived from both the metrics and associated visualizations, including loss plots and performance comparisons across different model configurations.

### 6.1 Performance of MLP Model on Actual Dataset

The baseline MLP model trained on the entire dataset without any drift awareness yields promising results with an MAE of 0.2438, RMSE of 0.3020, and R² of 0.8835. As shown in Figure 4, the model can capture the general consumption trends, but it fails to adapt effectively to distributional shifts over time. The training loss curve (Figure 5) confirms a steady convergence, but the relatively higher test error suggests suboptimal generalization post-drift.

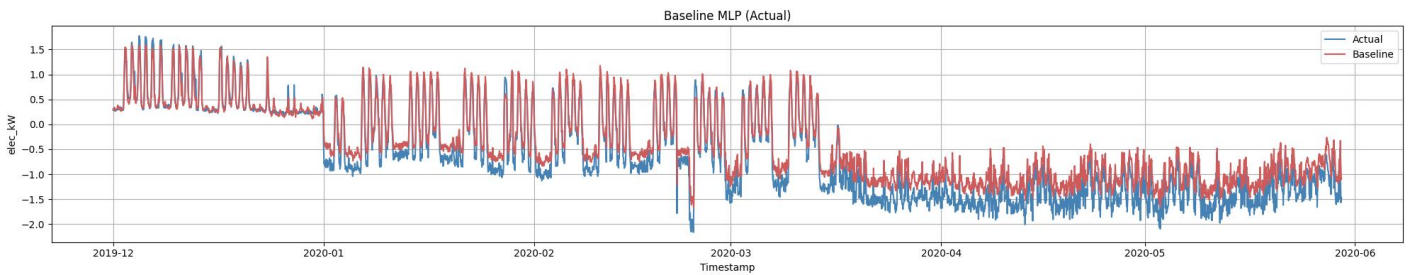

Figure 4: Actual vs Baseline MLP (Actual Dataset)

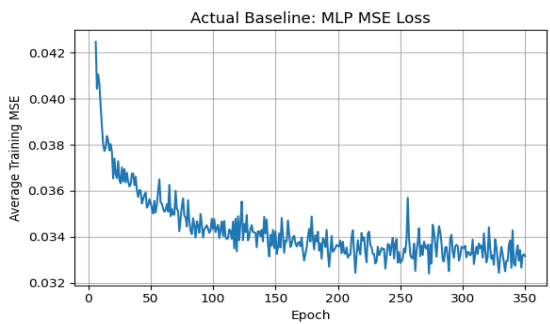
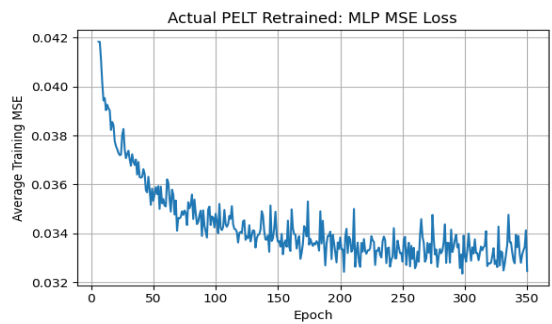

Figure 5: Actual MLP Baseline Loss                    Figure 6: Actual MLP Retrain Loss

In contrast, the PELT-retrained MLP model demonstrates significant improvement after retraining only on the drift-affected segment. With a reduced MAE of 0.1777, RMSE of 0.2454, and a notably higher R² of 0.9231, the retrained model better aligns with the true dynamics, especially during the most volatile periods. This is evident in Figure 7, where prediction closely follows actual values even across abrupt shifts. The training loss in Figure6 also reveals faster convergence, likely aided by training on a more stable post-drift distribution.

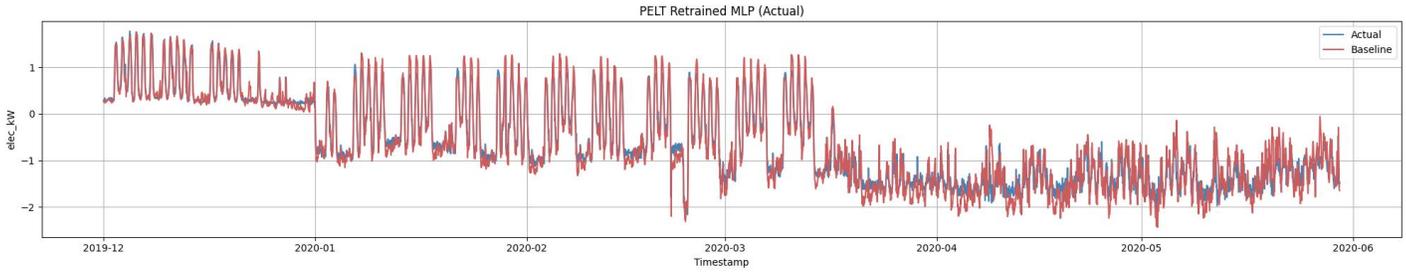

Figure 7: Actual vs Retrain MLP (Actual Dataset)

## 6.2 Performance of MLP Model on Synthetic Dataset

The synthetic dataset provides a controlled environment to rigorously assess the model's drift adaptability. The baseline MLP achieves MAE of 0.2931, RMSE of 0.3008, and $R^2$ of 0.877, as shown in Figure8. Although loss convergence is stable(Figure9), the baseline model cannot adjust to the manually injected drift regimes, resulting in prediction lag and mismatch.

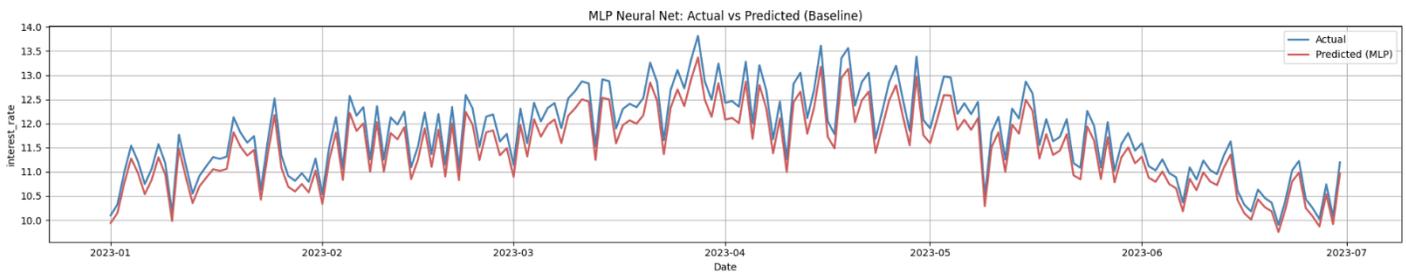

Figure 8: Actual vs Baseline MLP (Synthetic Dataset)

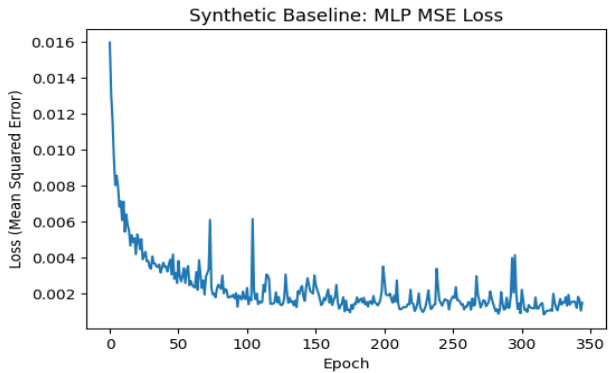
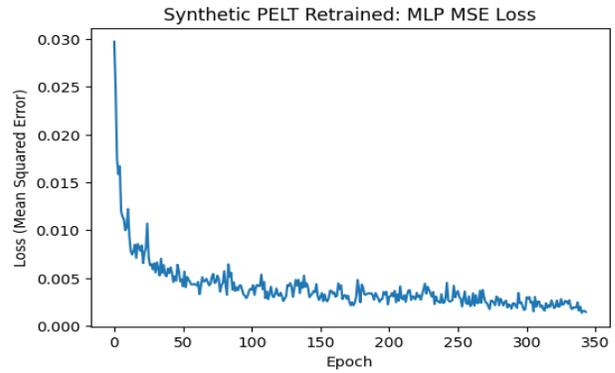

Figure 9: Synthetic MLP Baseline Loss    Figure 10: Synthetic MLP Retrain Loss

The PELT-retrained MLP, on the other hand, exhibits strong adaptation to detected drifts. After retraining on the segment post the last change point, it achieves MAE of 0.1636, RMSE of 0.1948, and a significantly enhanced $R^2$ of 0.9484. The predictive alignment in Figure11 is visually much tighter, while the loss curve (Figure10) reaffirms model stability even in retrained configurations.

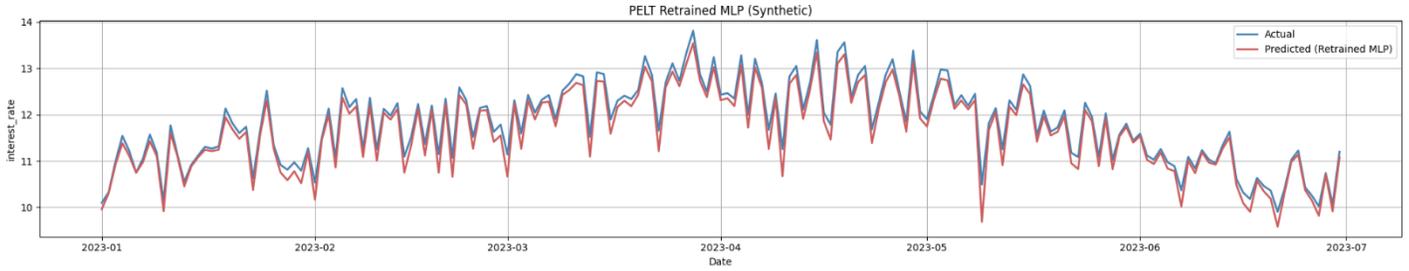
**Figure 11: Actual vs Retrain MLP (Synthetic Dataset)**

## 6.3 Lasso Performance on Actual Dataset

On the actual dataset, the baseline Lasso model yields an MAE of 0.4182, RMSE of 0.4984, and R² of 0.6828, indicating moderate fit quality under drift conditions (Figure12). Despite regularization via α=0.01, the model exhibits visible underfitting in drift-affected periods. Figure13 shows elevated error values, highlighting the baseline's limited robustness to shifting data regimes.

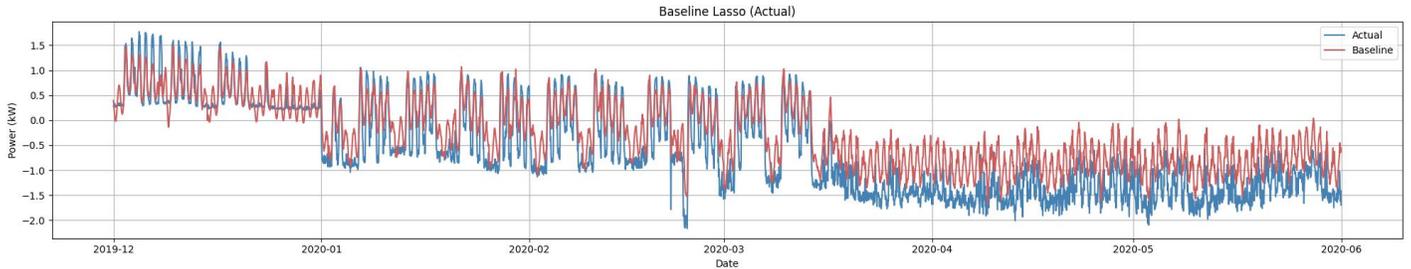
**Figure 12: Actual vs Baseline Lasso (Actual Dataset)**

Following PELT-triggered retraining on post-drift segments, the Lasso model shows substantial improvement, achieving an MAE of 0.1346, RMSE of 0.1897, and R² of 0.9540. The prediction alignment in Figure13 is visibly tighter, while the test MSE trend in Figure12 exhibits lower generalization error across regularization strengths, validating the effectiveness of selective retraining.

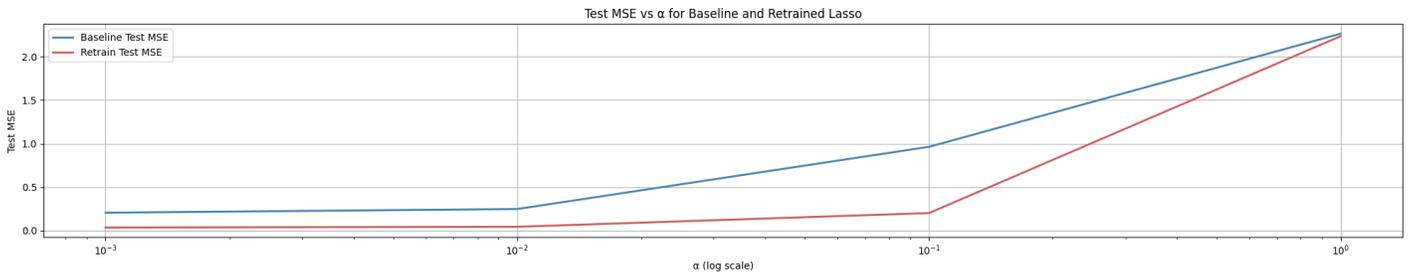
**Figure 13: Test MSE vs α for Baseline vs Retrain Lasso (Actual Dataset)**

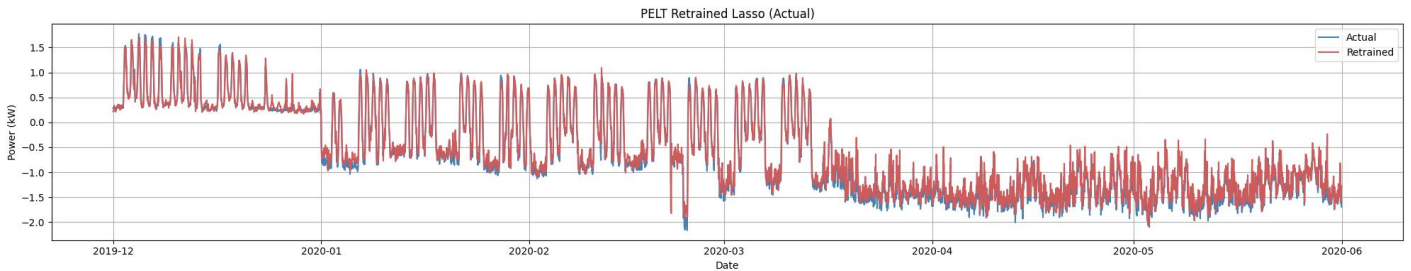
**Figure 14: Actual vs Retrain Lasso (Actual Dataset)**

## 6.4 Lasso Performance on Synthetic Dataset

The synthetic dataset allows us to evaluate the Lasso model's performance in a scenario with explicitly injected drifts and a clean feature space. The baseline Lasso model, trained globally without drift-awareness, achieves an MAE of 0.3357, RMSE of 0.4390, and R² of 0.7379 (Figure 15). While reasonable, the model struggles to generalize across drift regions, evident from its relatively high error and limited R². This underlines the inadequacy of a static regression model in the presence of temporal regime shifts.

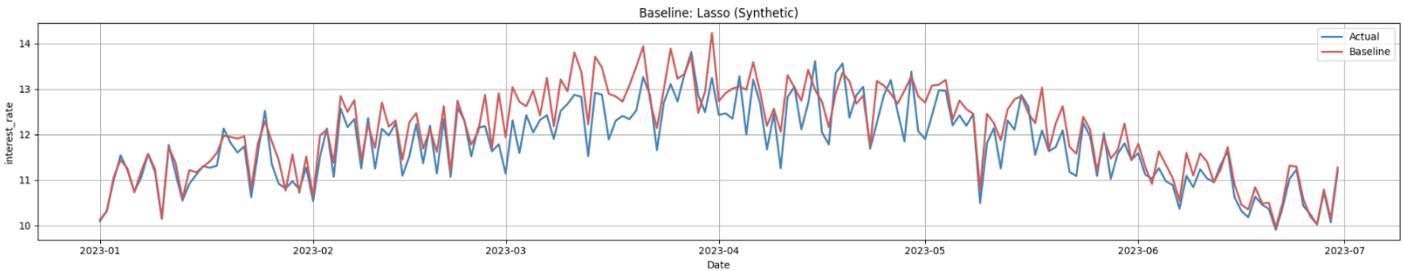

Figure 15: Actual vs Baseline Lasso (Synthetic Dataset)

To address this, PELT-based retraining is employed on the post-drift segment using polynomial feature expansion, which improves model expressiveness and adaptability. The retrained model yields a lower MAE of 0.2284, RMSE of 0.2882, and an improved R² of 0.8871, signaling stronger alignment with ground truth dynamics (Figure 16). Figure 17 further emphasizes that the retrained model generalizes better across a wider range of regularization strengths, showcasing improved adaptability through localized retraining.

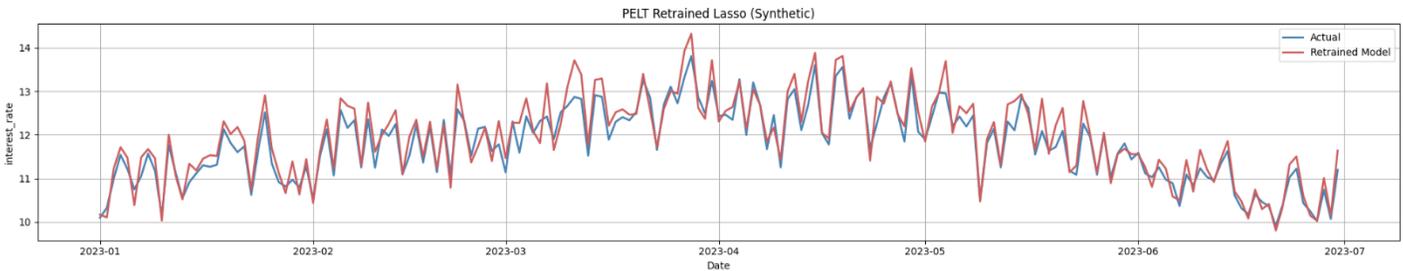

Figure 16: Actual vs Retrain Lasso (Synthetic Dataset)

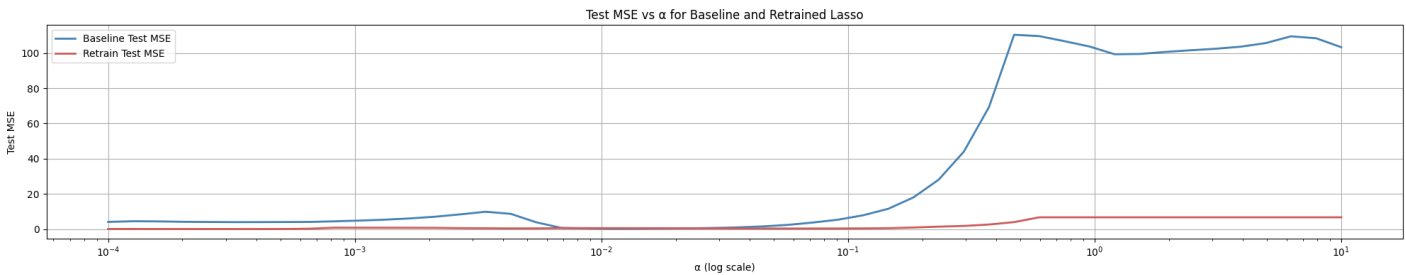

Figure 17: Test MSE vs α for Baseline vs Retrain Lasso (Synthetic Dataset)

## 6.5 Cross-Model and Dataset Performance Analysis

To unify the results discussed across individual experiments, Figure 18 presents a metric-wise visualization of MAE, RMSE, and R² for all models evaluated on both datasets. This design offers a clear separation of each performance metric, allowing us to analyze model behavior in a more focused and meaningful way. The top subplot, comparing MAE values, reveals that the PELT-retrained MLP model on the actual dataset achieves the lowest absolute error across all configurations. This indicates its superior capacity to adapt to evolving data patterns. Similarly, LassoCV retrained on the actual dataset also shows a substantial MAE drop compared to its baseline, validating that even linear models benefit from drift-aware retraining when properly tuned. The RMSE subplot further emphasizes this distinction. The retrained MLP again leads on the actual dataset with the lowest root mean squared error, reflecting its ability to reduce large prediction deviations. Interestingly, Lasso retrained on the synthetic dataset achieves competitive RMSE, almost matching MLP, which suggests that in lower-noise environments with structured drift, LassoCV especially with polynomial augmentation can act as a viable lightweight alternative to deep models.

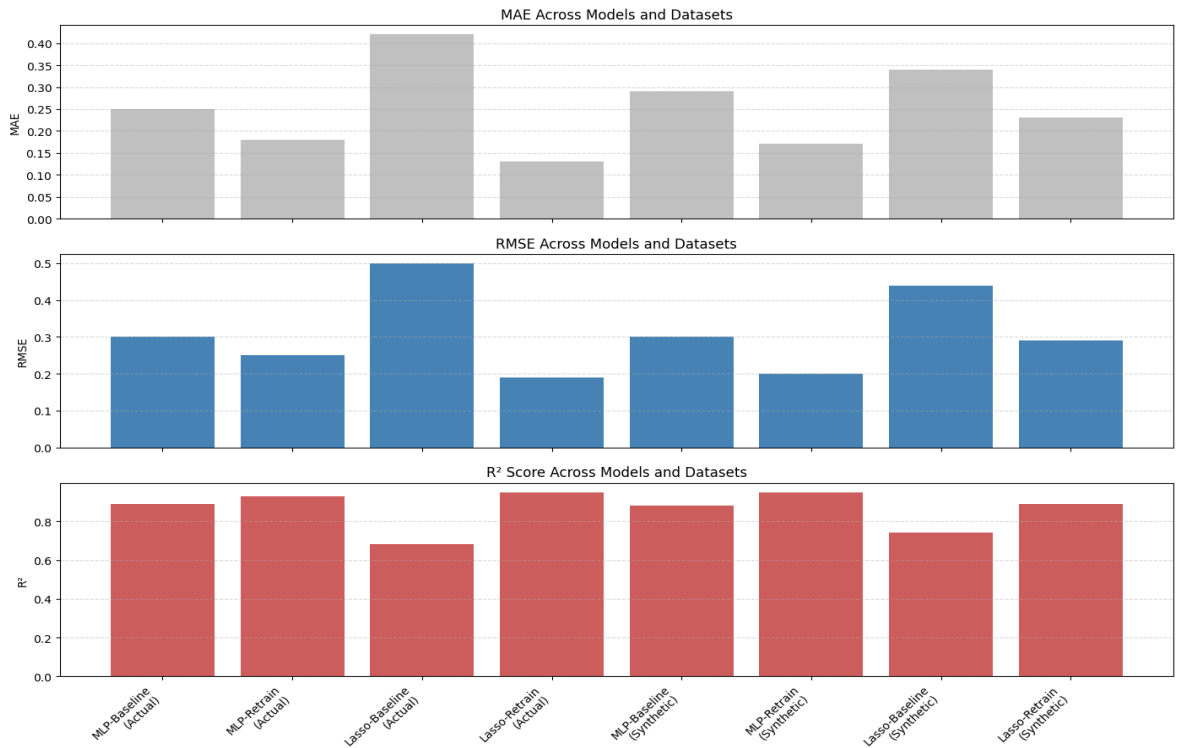

Figure 18: Cross-Model Metrics Comparison Across Datasets

In the R² score subplot, which quantifies how well the models explain variance in the target, retrained versions of both MLP and Lasso consistently outperform their respective baselines. The highest R² is achieved by the retrained MLP on the synthetic dataset, closely followed by the retrained Lasso on the actual dataset. This confirms that retraining after drift detection improves generalization across both complex and simpler domains.

Overall, Figure18 highlights that PELT-driven retraining yields consistent improvements in all three-performance metrics for both model families. MLP models demonstrate stronger gains in high-variability, real-world scenarios, while LassoCV remains a strong contender under synthetic conditions with structured changes. This separation of metrics also allows practitioners to choose models based on whether minimizing large errors (RMSE), general fit (R²), or overall deviation (MAE) is most critical for their specific application.

# 7 Discussion and Future Work

This study demonstrates that drift-aware retraining consistently improves forecasting performance across both real-world and synthetic datasets. As shown in Figure18, the retrained models both MLP and LassoCV achieve lower MAE and RMSE, alongside higher R² values, when compared to their baseline counterparts. These improvements are especially pronounced in the synthetic dataset, where the introduced distribution shifts allowed for a rigorous test of model adaptability. In particular, the synthetic dataset exhibits significant benefits from retraining, especially in the case of MLP. As visualized in Figure11, the retrained MLP aligns closely with the actual trend even during sharp fluctuations, unlike its baseline version. This improved alignment is supported by the loss convergence curve in Figure10, which demonstrates more stable and faster convergence after retraining. Similarly, the LassoCV model, when retrained on post-drift data with extended causal features, shows enhanced responsiveness to temporal and seasonal variations, reflected in its corresponding predictive alignment plot.

The actual dataset, while not exhibiting synthetic-level drift, still benefits from retraining. Figure7 reveals subtle yet meaningful improvements, where the retrained model demonstrates a tighter fit, especially during seasonal peaks. LassoCV's retrained variant also captures evolving patterns more effectively, as seen in Figure14. These visual insights corroborate the quantitative metrics, even moderate real-world drift, if unaddressed, can degrade model generalization. A critical factor in the success of the retraining strategy lies in the integration of causal and trend-aware features post-drift. These features such as rolling means, lagged differences, and harmonic components enhance temporal sensitivity, particularly when used with dynamic model retraining triggered by the PELT change-point detection algorithm. Across both datasets and model architectures, the results validate our central hypothesis proactively adapting models to distributional changes substantially boost forecasting accuracy and resilience. Furthermore, the consistent improvements across both linear (LassoCV) and nonlinear (MLP) models underscore the model-agnostic nature of our approach. This strengthens the argument that the underlying drift-adaptive methodology rather than the specific choice of model is the key contributor to the observed gains.

Building on the findings of this research, several promising avenues for future exploration are identified. A critical next step involves deploying the pipeline on real-time embedded systems such as IoT devices, smart meters, or autonomous platforms, where lightweight, low-latency drift adaptation becomes essential for real-world applicability. Transitioning from batch-mode processing to real-time deployment would involve optimizing changepoint detection and retraining pipelines to operate within the strict memory, power, and latency constraints of embedded environments. This effort bridges the current research with practical use cases in operational settings. In addition, expanding the forecasting horizon remains a valuable direction. While this study focuses on short-term prediction, evaluating model robustness across hourly, daily, and weekly horizons would provide insights into temporal generalizability. It would also allow for tailored retraining strategies based on different application requirements, shorter horizons may benefit from rapid adaptation, while longer horizons demand structural stability and trend awareness. This exploration could directly impact domains like energy load balancing, demand planning, or climate-sensitive forecasting where time scale matters. Together, these future directions emphasize the importance of scalability and temporal adaptability in ensuring that drift-aware predictive models remain robust and usable beyond controlled experimental conditions.

## 8 Conclusion

This study introduces a robust and interpretable framework for addressing concept drift in time series regression through changepoint-aware retraining. By leveraging PELT-based drift detection and applying targeted model retraining after the last identified drift, we observed marked improvements in forecast reliability across both synthetic and real-world datasets. The use of two fundamentally distinct models, MLP for non-linear learning and LassoCV for sparse linear regression allowed us to validate the consistency and generalizability of the approach across varied modeling paradigms. Our findings underscore the importance of adapting to temporal distribution shifts rather than relying on static models. The retrained models showed more accurate alignment with recent data patterns, capturing both abrupt and gradual changes in the underlying signal. Notably, this performance gain was achieved without altering model architecture or tuning, reinforcing the effectiveness of retraining as a standalone strategy. By designing a controlled experimental setup and applying identical feature engineering across models, we isolated drift adaptation as the primary driver of performance improvement. This methodological rigor strengthens the reliability of our conclusions and provides a replicable path for future studies. As time-dependent predictive systems become central to sectors like energy forecasting, retail demand, and transportation planning, our work presents a practical, scalable solution to combat concept drift. It opens pathways toward resilient, drift-adaptive forecasting pipelines capable of sustaining accuracy in dynamically changing environments.

**Author Contributions:** Conceptualization, V.-H.B. and N.P.; methodology, N.P., W.S., G.H.; soft- ware, N.P. and G.H.; validation; A.H. and G.H.; formal analysis, N.P. and A.H.; resources, N.P. and A.H.; data curation, N.P; writing—original draft preparation, N.P.; writing—review and editing, V.-H.B., W.S., G.H.; visualization, N.P. and A.H.; supervision, V.-H.B., W.S.; project administration, V.-H.B.; funding acquisition, V.-H.B. All authors have read and agreed to the published version of the manuscript.

**Funding:** The author's work was supported by the University of Michigan-Dearborn's Office of Research "Research Initiation & Development".

**Institutional Review Board Statement:** Not applicable.
**Informed Consent Statement:** Not applicable.
**Data Availability Statement:** https://datadryad.org/dataset/doi:10.7941/D1N33Q
**Conflicts of Interest:** The authors declare no conflicts of interest

# References

[1] Zhao, L., & Shen, Y. (2024). Proactive Model Adaptation Against Concept Drift for Online Time Series Forecasting. *arXiv preprint arXiv:2412.08435*.

[2] Liu, J., Yang, D., Zhang, K., Gao, H., & Li, J. (2023). Anomaly and Change Point Detection for Time Series with Concept Drift. *World Wide Web*, 26(6), 3229–3252.

[3] Bayram, F., Aupke, P., Ahmed, B. S., Kassler, A., Theocharis, A., & Forsman, J. (2023). DA-LSTM: A Dynamic Drift-Adaptive Learning Framework for Interval Load Forecasting with LSTM Networks. *Engineering Applications of Artificial Intelligence*, 123, 106480.

[4] Liu, Z., Godahewa, R., Bandara, K., & Bergmeir, C. (2023). Handling Concept Drift in Global Time Series Forecasting. *arXiv preprint arXiv:2304.01512*.

[5] He, H., Zhang, Q., Yi, K., Shi, K., Niu, Z., & Cao, L. (2022). Distributional Drift Adaptation with Temporal Conditional Variational Autoencoder for Multivariate Time Series Forecasting. *arXiv preprint arXiv:2209.00654*.

[6] Oliveira, D. R., Oliveira, A. L. I., & Batista, G. E. A. P. A. (2017). Time Series Forecasting in the Presence of Concept Drift: A PSO-Based Approach. *IEEE 29th International Conference on Tools with Artificial Intelligence (ICTAI)*, 1, 1–8.

[7] Souza, V. M. A., Reis, D. M., Maletzke, A. G., & Batista, G. E. A. P. A. (2020). Challenges in Benchmarking Stream Learning Algorithms with Real-World Data. *Data Mining and Knowledge Discovery*, 34(2), 409–439.

[8] Gama, J., Žliobaitė, I., Bifet, A., Pechenizkiy, M., & Bouchachia, A. (2014). A Survey on Concept Drift Adaptation. *ACM Computing Surveys*, 46(4), 1–37.

[9] Widmer, G., & Kubat, M. (1996). Learning in the Presence of Concept Drift and Hidden Contexts. *Machine Learning*, 23(1), 69–101.

[10] Lu, J., Liu, A., Dong, F., Gu, F., & Gama, J. (2018). Learning Under Concept Drift: A Review. *IEEE Transactions on Knowledge and Data Engineering*, 31(12), 2346–2363.

[11] Gomes, H. M., Read, J., Bifet, A., Barddal, J. P., & Gama, J. (2017). Adaptive Random Forests for Evolving Data Stream Classification. *Machine Learning*, 106(9-10), 1469–1495.

[12] Žliobaitė, I., Bifet, A., Read, J., Pfahringer, B., & Holmes, G. (2015). Evaluation Methods and Decision Theory for Classification of Streaming Data with Temporal Dependence. *Machine Learning*, 98(3), 455–482.

[13] Alippi, C., & Polikar, R. (2014). Guest Editorial: Learning in Nonstationary and Evolving Environments. *IEEE Transactions on Neural Networks and Learning Systems*, 25(1), 1–4.

[14] Basseville, M., & Nikiforov, I. V. (1993). Detection of Abrupt Changes: Theory and Application. *Prentice-Hall*.

[15] Krawczyk, B., Minku, L. L., Gama, J., Stefanowski, J., & Wozniak, M. (2017). Ensemble Learning for Data Stream Analysis: A Survey. *Information Fusion*, 37, 132–156.

[16] Killick, R., Fearnhead, P., & Eckley, I. A. (2012). Optimal detection of changepoints with a linear computational cost. *Journal of the American Statistical Association*, *107*(500), 1590-1598.

[17] Yang, W., Li, Z., Liu, M., Lu, Y., Cao, K., Maciejewski, R., & Liu, S. (2020, October). Diagnosing concept drift with visual analytics. In *2020 IEEE conference on visual analytics science and technology (VAST)* (pp. 12-23). IEEE.

[18] Hadry, M., Grillmeyer, D., Becker, M., König, M., Lesch, V., & Kounev, S. (2024, September). Evaluating the Benefits of Model Retraining for Self-Aware Vehicle Traffic Forecasting. In *2024 IEEE International Conference on Autonomic Computing and Self-Organizing Systems (ACSOS)* (pp. 41-50). IEEE.

[19] Bayram, F., Ahmed, B. S., & Kassler, A. (2022). From concept drift to model degradation: An overview on performance-aware drift detectors. *Knowledge-Based Systems*, *245*, 108632.

[20] Chai, T., & Draxler, R. R. (2014). Root mean square error (RMSE) or mean absolute error (MAE)?–Arguments against avoiding RMSE in the literature. *Geoscientific model development*, *7*(3), 1247-1250.